\documentclass[journal]{IEEEtran}
\usepackage{booktabs}
\usepackage{hhline}
\usepackage{multirow}
\usepackage{capt-of}
\usepackage{listings}
\usepackage{colortbl}
\usepackage{todonotes}

\makeatletter

 \AtBeginDocument{%
  \providecommand*\ext@lstlisting{lol}%
  \renewcommand{\fnum@lstlisting}{\lstlistingname\nobreakspace\thelstlisting}
 }
\makeatother

\usepackage{amsmath}
\usepackage{dsfont}
\newcommand\floor[1]{\lfloor#1\rfloor}
\usepackage{algpseudocode}

\begin{document}
%
\title{Attention-Based Neural Networks for Chroma Intra Prediction in Video Coding}
%
%
%

\author{Marc~Górriz~Blanch,~\IEEEmembership{Student~Member~IEEE,}
        Saverio~Blasi,
        Alan~F.~Smeaton,~\IEEEmembership{Fellow~IEEE,}
        Noel~E.~O’Connor,~\IEEEmembership{Member~IEEE,}
        and~Marta~Mrak,~\IEEEmembership{Senior Member~IEEE}
\thanks{Manuscript submitted July 1, 2020. The work described in this paper has been conducted within the project JOLT funded by the European Union’s Horizon 2020 research and innovation programme under the Marie Skłodowska Curie grant agreement No 765140.}
\thanks{M. Górriz Blanch, S. Blasi and M. Mrak are with BBC Research \& Development, The Lighthouse, White City Place, 201 Wood Lane, London, UK (e-mail: marc.gorrizblanch@bbc.co.uk, saverio.blasi@bbc.co.uk, marta.mrak@bbc.co.uk).}
\thanks{A. F. Smeaton and N. E. O'Connor are with Dublin City University, Glasnevin, Dublin 9, Ireland (e-mail: alan.smeaton@dcu.ie, noel.oconnor@dcu.ie).}
}

%
%

\markboth{Journal of Selected Topics in Signal Processing, October~2020}%
{Journal of Selected Topics in Signal Processing, October~2020}
%



\maketitle

\begin{abstract}
Neural networks can be successfully used to improve several modules of advanced video coding schemes. In particular, compression of colour components was shown to greatly benefit from usage of machine learning models, thanks to the design of appropriate attention-based architectures that allow the prediction to exploit specific samples in the reference region. However, such architectures tend to be complex and computationally intense, and may be difficult to deploy in a practical video coding pipeline. This work focuses on reducing the complexity of such methodologies, to design a set of simplified and cost-effective attention-based architectures for chroma intra-prediction. A novel size-agnostic multi-model approach is proposed to reduce the complexity of the inference process. The resulting simplified architecture is still capable of outperforming state-of-the-art methods. Moreover, a collection of simplifications is presented in this paper, to further reduce the complexity overhead of the proposed prediction architecture. Thanks to these simplifications, a reduction in the number of parameters of around 90\% is achieved with respect to the original attention-based methodologies. Simplifications include a framework  for  reducing the overhead of  the  convolutional operations, a simplified cross-component processing model integrated into the original architecture, and a methodology to perform integer-precision approximations with the aim to obtain fast and hardware-aware implementations. The proposed schemes are integrated into the Versatile Video Coding (VVC) prediction pipeline, retaining compression efficiency of state-of-the-art chroma intra-prediction methods based on neural networks, while offering different directions for significantly reducing coding complexity.
\end{abstract}

\begin{IEEEkeywords}
Chroma intra prediction, convolutional neural networks, attention algorithms, multi-model architectures, complexity reduction, video coding standards.
\end{IEEEkeywords}

%
\IEEEpeerreviewmaketitle

\section{Introduction}
%
%
%
%

\begin{figure}[!t]
\centering
\includegraphics[width=0.42\textwidth]{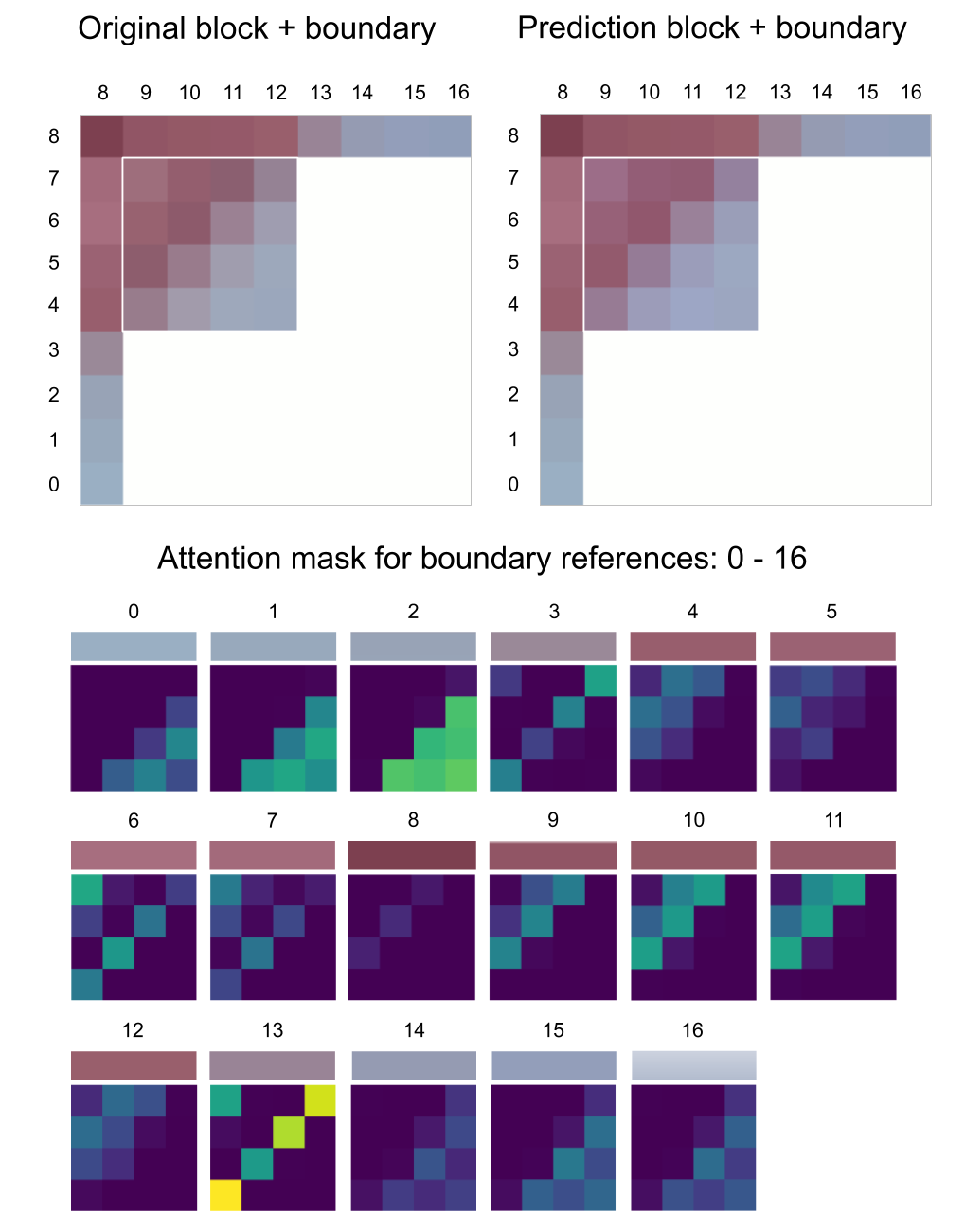}
\caption{Visualisation of the attentive prediction process. For each reference sample 0-16 the attention module generates its contribution to the prediction of individual pixels from a target $4 \times 4$ block.}
\label{att_interpret}
\end{figure}

\IEEEPARstart{E}{fficient} video compression has become an essential component of multimedia streaming. The convergence of digital entertainment followed by the growth of web services such as video conferencing, cloud gaming and real-time high-quality video streaming, prompted the development of advanced video coding technologies capable of tackling the increasing demand for higher quality video content and its consumption on multiple devices. New compression techniques enable a compact representation of video data by identifying and removing spatial-temporal and statistical redundancies within the signal. This results in smaller bitstreams, enabling more efficient storage and transmission as well as distribution of content at higher quality, requiring reduced resources. 

Advanced video compression algorithms are often complex and computationally intense, significantly increasing  the encoding and decoding time. Therefore, despite bringing high coding gains, their potential for application in practice is limited.
Among the current state-of-the-art solutions, the next generation Versatile Video Coding standard \cite{bross2019vvcdraft} (referred to as VVC in the rest of this paper), targets between 30-50\% better compression rates for the same perceptual quality, supporting resolutions from 4K to 16K as well as 360$^{\circ}$ videos. 
One fundamental component of hybrid video coding schemes, intra prediction, exploits spatial redundancies within a frame by predicting samples of the current block from already reconstructed samples in its close surroundings. VVC allows a large number of possible intra prediction modes to be used on the luma component at the cost of a considerable amount of signalling data. Conversely, to limit the impact of mode signalling, chroma components employ a reduced set of modes \cite{bross2019vvcdraft}.

In addition to traditional modes, more recent research introduced schemes which further exploit cross-component correlations between the luma and chroma components. Such correlations motivated the development of the Cross-Component Linear Model (CCLM, or simply LM in this paper) intra modes. When using CCLM, the chroma components are predicted from already reconstructed luma samples using a linear model. Nonetheless, the limitation of simple linear predictions comes from its high dependency on the selection of predefined reference samples. Improved performance can be achieved using more sophisticated Machine Learning (ML) mechanisms \cite{li2018hybrid, pfaff2018intra}, which are able to derive more complex representations of the reference data and hence boost the prediction capabilities.

Methods based on Convolutional Neural Networks (CNNs) \cite{li2018hybrid, grriz2020chroma} provided significant improvements at the cost of two main drawbacks: the associated increase in system complexity and the tendency to disregard the location of individual reference samples. Related works deployed complex neural networks (NNs) by means of model-based interpretability \cite{murn2020interpreting}. For instance, VVC recently adopted simplified NN-based methods such as Matrix Intra Prediction (MIP) modes \cite{helle2019intra} and Low-Frequency Non Separable Transform (LFNST) \cite{zhao2016nsst}. For the particular task of block-based intra-prediction, the usage of complex NN models can be counterproductive if there is no control over the relative position of the reference samples. When using fully-connected layers, all input samples contribute to all output positions, and after the consecutive application of several hidden layers, the location of each input sample is lost. This behaviour clearly runs counter to the design of traditional approaches, in which predefined directional modes carefully specify which boundary locations contribute to each prediction position. A novel ML-based cross-component intra-prediction method is proposed in \cite{grriz2020chroma}, introducing a new attention module capable of tracking the contribution of each neighbouring reference sample when computing the prediction of each chroma pixel, as shown in Figure \ref{att_interpret}. As a result, the proposed scheme better captures the relationship between the luma and chroma components, resulting in more accurate prediction samples. However, such NN-based methods significantly increase the codec complexity, increasing the  encoder  and  decoder  times by  up  to 120\% and 947\%, respectively.

This paper focuses on complexity reduction in video coding with the aim to derive a set of simplified and cost-effective attention-based architectures for chroma intra-prediction. Understanding and distilling  knowledge from the networks enables the implementation of less complex algorithms which achieve similar performance to the original models. Moreover, a novel training methodology is proposed in order to design a block-independent multi-model which outperforms the state-of-the-art attention-based architectures and reduces inference complexity. The use of variable block sizes during training helps the model to better generalise on content variety while ensuring higher precision on predicting large chroma blocks. The main contributions of this work are the following:

\begin{itemize}
    \item A competitive block-independent attention-based multi-model and training methodology;
    \item A framework for complexity reduction of the convolutional operations;
    \item A simplified cross-component processing model using sparse auto-encoders;
    \item A fast and cost-effective attention-based multi-model with integer precision approximations.
\end{itemize}

This paper is organised as follows: Section~\ref{sec:background} provides a brief overview on the related work, Section~\ref{sec:attention} introduces the attention-based methodology in detail and establishes the mathematical notation for the rest of the paper, Section~\ref{sec:simplifications} presents the proposed simplifications and Section~\ref{sec:experiments} shows experimental results, with conclusion drawn in Section~\ref{sec:conclusions}.

\begin{figure*}[!t]
\centering
\includegraphics[width=0.70\textwidth]{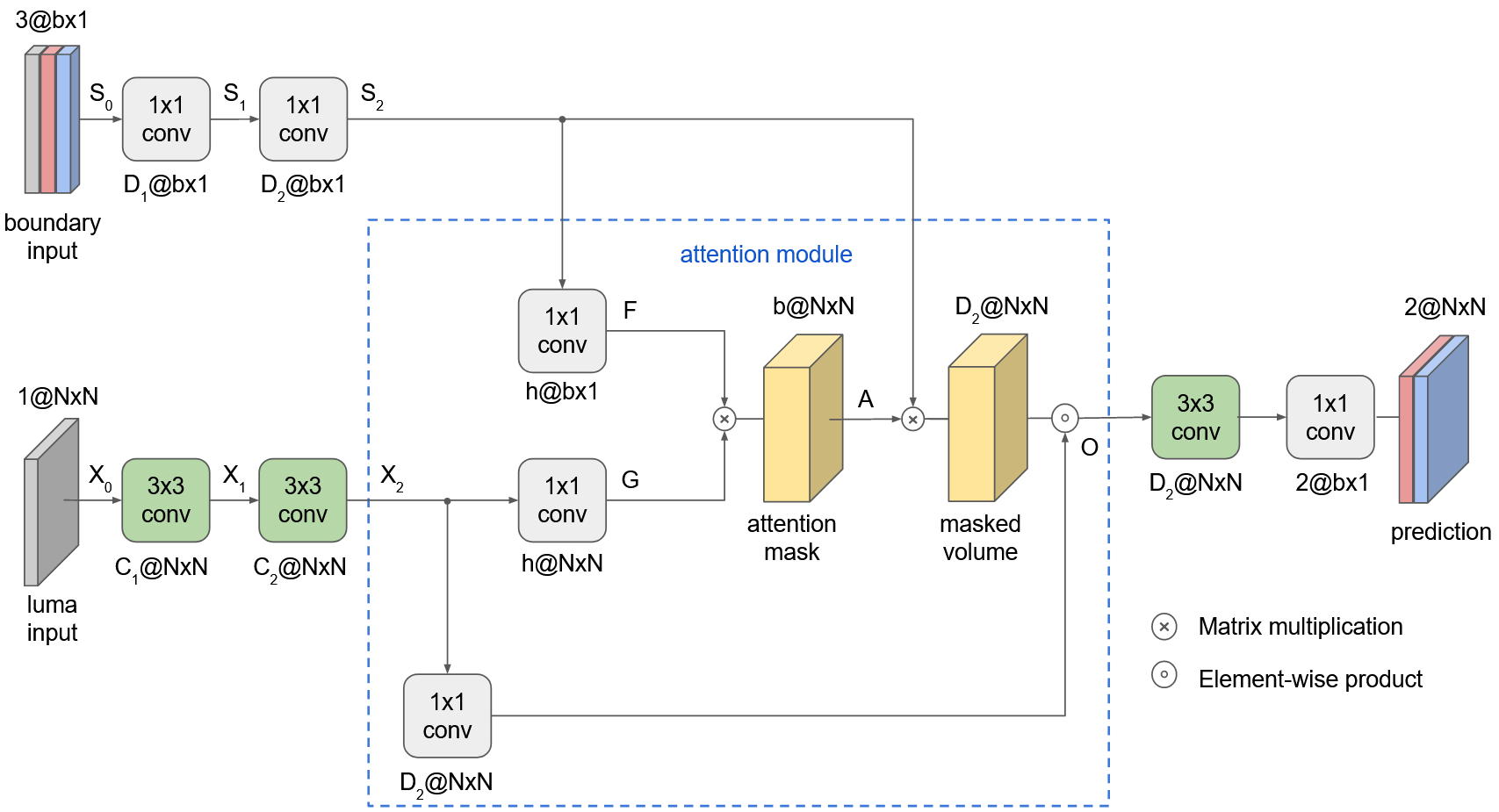}
\caption{Baseline attention-based architecture for chroma intra prediction presented in \cite{grriz2020chroma} and described in Section \ref{sec:attention}.}
\label{icip-arch}
\end{figure*}

\section{Background}
\label{sec:background}

Colour images are typically represented by three colour components (e.g. RGB, YCbCr). The YCbCr colour scheme is often adopted by digital image and video coding standards (such as JPEG, MPEG-1/2/4 and H.261/3/4) due to its ability to compact the signal energy and to reduce the total required bandwidth. Moreover, chrominance components are often subsampled by a factor of two to conform to the YCbCr 4:2:0 chroma format, in which the luminance signal contains most of the spatial information. Nevertheless,  cross-component redundancies can be further exploited by reusing information from already coded components to compress another component. In the case of YCbCr, the Cross-Component Linear model (CCLM) \cite{zhang2018enhanced} uses a linear model to predict the chroma signal from a subsampled version of the already reconstructed luma block signal. The model parameters are derived at both the encoder and decoder sides without needing explicit signalling in the bitstream. 

Another example is the Cross-Component Prediction (CCP) \cite{nguyen2013adaptive} which resides at the transform unit (TU) level regardless of the input colour space. In case of YCbCr, a subsampled and dequantised luma transform block (TB) is used to modify the chroma TB at the same spatial location based on a context parameter signalled in the bitstream. An extension of this concept modifies one chroma component using the residual signal of the other one \cite{siekmann2017extended}. Such methodologies significantly improved the coding efficiency by further exploiting the cross-component correlations within the chroma components.

In parallel, recent success of deep learning application in computer vision and image processing influenced design of novel video compression algorithms. In particular in the context of intra-prediction, a new algorithm \cite{pfaff2018intra}  was introduced based on fully-connected layers and CNNs to map the prediction of block positions from the already reconstructed neighbouring samples, achieving BD-rate (Bjontegaard Delta rate) \cite{bjontegaard2001calculation} savings of up to 3.0\% on average over HEVC, for approx. 200\% increase in decoding time. The successful integration of CNN-based methods for luma intra-prediction into existing codec architectures has motivated research into alternative methods for chroma prediction, exploiting cross-component redundancies similar to the aforementioned LM methods. A novel hybrid neural network for chroma intra prediction was recently introduced in \cite{li2018hybrid}. A first CNN was designed to extract features from reconstructed luma samples. This was combined with another fully-connected network used to extract cross-component correlations between neighbouring luma and chroma samples. The resulting architecture uses complex non-linear mapping for end-to-end prediction of chroma channels. However, this is achieved at the cost of disregarding the spatial location of the boundary reference samples and significant increase of the complexity of the prediction process. As shown in \cite{grriz2020chroma}, after a consecutive application of fully-connected layers in \cite{li2018hybrid}, the location of each input boundary reference sample is lost. Therefore, the fully-convolutional architecture in \cite{grriz2020chroma} better matches the design of the directional VVC modes and is able to provide significantly better performance.

The use of attention models enables effective utilisation of the individual spatial location of the reference samples \cite{grriz2020chroma}. The concept of ``attention-based” learning is a well-known idea used in deep learning frameworks, to improve the performance of trained networks in complex prediction tasks \cite{vaswani2017attention, lin2017structured, parikh2016decomposable}. In particular, self-attention is used to assess the impact of particular input variables on the outputs, whereby the prediction is computed focusing on the most relevant elements of the same sequence \cite{cheng2016long}. The novel attention-based architecture introduced in \cite{grriz2020chroma} reports average BD-rate reductions of -0.22\%, -1.84\% and -1.78\% for the Y, Cb and Cr components, respectively, although it significantly impacts the encoder and decoder time. 

One common aspect across all related work is that whilst the result is an improvement in compression this comes at the expense of increased complexity of the encoder and decoder. In order to address the complexity challenge, this paper aims to design a set of simplified attention-based architectures for performing chroma intra-prediction faster and more efficiently. Recent works addressed complexity reduction in neural networks using methods such as channel pruning \cite{he2017channel, zhuang2018discrimination, chin2020towards} and quantisation \cite{jacob2018quantization, cai2020zeroq, xu2020generative}. In particular for video compression, many works used integer arithmetic in order to efficiently implement trained neural networks on different hardware platforms. For example, the work in \cite{courbariaux2014training} proposes a training methodology to handle low precision multiplications, proving that very low precision is sufficient not just for running trained networks but also for training them. Similarly, the work in \cite{balle2018integer} considers the problem of using variational latent-variable models for data compression and proposes integer networks as a universal solution of range coding as an entropy coding technique. They demonstrate that such models enable reliable cross-platform encoding and decoding of images using variational models. Moreover, in order to ensure deterministic implementations on hardware platforms, they approximate non-linearities using lookup tables. Finally, an efficient implementation of matrix-based intra prediction is proposed in \cite{schafer2020efficient}, where a performance analysis evaluates the challenges of deploying models with integer arithmetic in video coding standards. Inspired by this knowledge, this paper develops a fast and cost-effective implementation of the proposed attention-based architecture using integer precision approximations. As shown Section~\ref{subsec:simulation}, while such approximations can significantly reduce the complexity, the associated drop of performance is still not negligible.

\begin{figure*}[!t]
\centering
\includegraphics[width=1\textwidth]{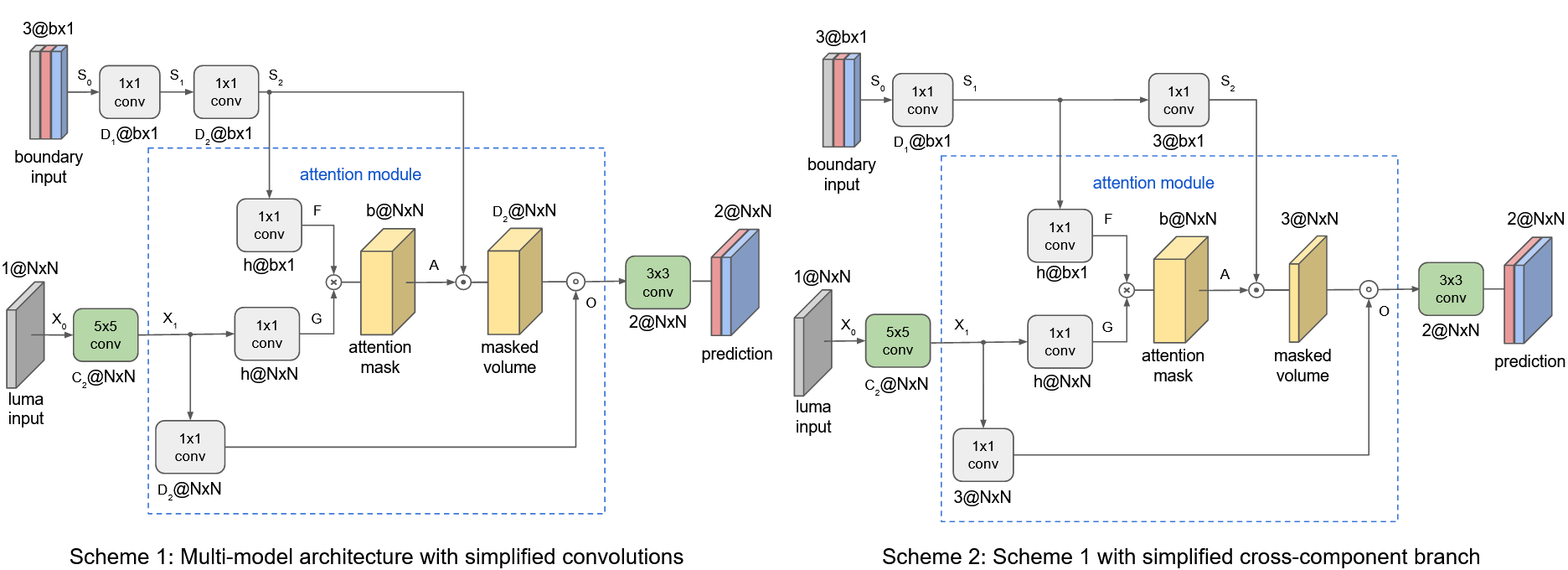}
\caption{Proposed multi-model attention-based architectures with the integration of the simplifications introduced in this paper. More details about the model's hyperparameters and a description of the referred schemes can be found in Section \ref{sec:experiments}.}
\label{schemes}
\end{figure*}

\section{Attention-based architectures}
\label{sec:attention}

This section describes in detail the attention-based approach proposed in \cite{grriz2020chroma} (Figure~\ref{icip-arch}), which will be the baseline for the presented methodology in this paper. The section also provides the mathematical notation used for the rest of this paper. 

Without loss of generality, only square blocks of pixels are considered in this work. After intra-prediction and reconstruction of a luma block in the video compression chain, luma samples can be used for prediction of co-located chroma components. In this discussion, the size of a luma block is assumed to be (downsampled to) $N \times N$ samples, which is the size of the co-located chroma block. This may require the usage of conventional downsampling operations, such as in the case of using chroma sub-sampled picture formats such as 4:2:0. Note that a video coding standard treats all image samples as unsigned integer values within a certain precision range based on the internal bit depth. However, in order to utilise common deep learning frameworks, all samples are converted to floating point and normalised to values within the range  $[0, 1]$. For the chroma prediction process, the  reference samples used include the co-located luma block $X_0\in {\rm I\!R}^{N \times N}$, and the array of reference samples $B_c\in {\rm I\!R}^{b}$, $b = 4N + 1$ from the left and from above the current block (Figure~\ref{att_interpret}), where $c = Y$, $C_{b}$ or $C_{r}$ refers to the three colour components. $B$ is constructed from  samples on the left boundary (starting from the bottom-most sample), then the corner is added, and finally the samples on top are added (starting from the left-most sample). In case some reference samples are not available, these are padded using a predefined value, following the standard approach defined in VVC. Finally, $S_0\in {\rm I\!R}^{3 \times b}$ is the cross-component volume obtained by concatenating the three reference arrays $B_{Y}$, $B_{Cb}$ and $B_{Cr}$. Similar to the model in \cite{li2018hybrid}, the attention-based architecture adopts a scheme based on three network branches that are combined to produce prediction samples, illustrated in Figure~\ref{icip-arch}. The first two branches work concurrently to extract features from the input reference samples. 

The first branch (referred to as the cross-component boundary branch) extracts cross component features from $S_0\in {\rm I\!R}^{3 \times b}$ by applying $I$ consecutive $D_{i}$~-~dimensional $1 \times 1$ convolutional layers to obtain the $S_{i}\in {\rm I\!R}^{D_{i} \times b}$ output feature maps, where $i = {1, 2 \dots I}$. By applying $1 \times 1$ convolutions, the boundary input dimensions are preserved, resulting in an $D_{i}$-dimensional vector of cross-component information for each boundary location. The resulting volumes are activated using a Rectified Linear Unit (ReLU) non-linear function.

In parallel, the second branch (referred to as the luma convolutional branch) extracts spatial patterns over the co-located reconstructed luma block $X_0$ by applying convolutional operations. The luma convolutional branch is defined by $J$ consecutive $C_{j}$-dimensional $3 \times 3$ convolutional layers with a stride of $1$, to obtain $X_{j}\in {\rm I\!R}^{C_{j} \times N^{2}}$ feature maps from the $N^{2}$ input samples, where $j = {1, 2 \dots J}$. Similar to the cross-component boundary branch, in this branch a bias and a ReLU activation are applied within convolutional layer.

The feature maps ($S_I$ and $X_J$) from both branches are each convolved using a $1 \times 1$ kernel, to project them into two corresponding reduced feature spaces. Specifically, $S_I$ is convolved with a filter $W_{F}\in {\rm I\!R}^{h \times D}$ to obtain the $h$-dimensional feature matrix $F$. Similarly, $X_J$ is convolved with a filter $W_{G}\in {\rm I\!R}^{h \times C}$ to obtain the $h$-dimensional feature matrix $G$. The two matrices are multiplied together to obtain the pre-attention map $M=G^{T} F$. Finally, the attention matrix $A\in {\rm I\!R}^{N^2 \times b}$ is obtained applying a softmax operation to each element of $M$, to generate the probability of each boundary location  being able to predict a sample location in the block. Each value $\alpha_{j, i}$ in $A$ is obtained as:
\begin{equation}
\alpha_{j, i} = \frac{\exp{(m_{i, j}/T})}{ \sum_{n=0}^{{b}-1} \exp{(m_{n, j}/T})},\label{eq1}
\end{equation}
\noindent where $j=0,...,N^{2}-1$ represents the sample location in the predicted block, $i=0,...,b-1$ represents a reference sample location, and $T$ is the softmax temperature parameter controlling the smoothness of the generated probabilities, with $0 < T \leq 1$. Notice that the smaller the value of $T$, the more localised are the obtained attention areas resulting in correspondingly fewer boundary samples contributing to a given prediction location. The weighted sum of the contribution of each reference sample in predicting a given sample at a specific location is obtained by computing the matrix multiplication between the cross-component boundary features $S_I$ and the attention matrix $A$, or formally $S_I^TA$. In order to further refine $S_I^TA$, this weighted sum can be multiplied by the output of the luma branch. To do so, the output of the luma branch must be transformed to change its dimensions by means of a $1 \times 1$ convolution using a matrix $W_{\bar{x}}\in {\rm I\!R}^{D \times C}$ to obtain a transformed representation $\bar{X}$, then $O = \bar{X} \odot (S_I^TA)$, where $\odot$ is the element-wise product.

Finally, the output of the attention model is  fed into the third network branch, to compute the predicted chroma samples. In this branch, a final CNN is used to map the fused features from the first two branches as combined by means of the attention model into the final chroma prediction. The prediction head branch is defined by two convolutional layers, applying $E$-dimensional $3 \times 3$ convolutional filters and then $2$-dimensional $1 \times 1$ filters for deriving the two chroma components at once.

\section{Multi-model architectures}
\label{sec:simplifications}

This section introduces a new multi-model architecture which improves the baseline attention-based approach (Section \ref{sec:attention}, \cite{grriz2020chroma}). The main improvement comes from its block-size agnostic property as the proposed approach only requires one model for all block sizes. Furthermore, a range of simplifications is proposed with the aim to reduce the complexity of related attention-based architectures while preserving prediction performance as much as possible. The proposed simplifications include a framework for complexity reduction of the convolutional operations, a simplified cross-component boundary branch using sparse autoencoders and insights for fast and cost-effective implementations with integer precision approximations. Figure \ref{schemes} illustrates the proposed multi-model attention-based schemes with the integration of the simplifications described in this section.

\subsection{Multi-model size agnostic architecture}
\label{subsec:multi-model}

In order to handle variable block sizes, previous NN-based chroma intra-prediction methods employ different architectures for blocks of different sizes. These architectures differ in the dimensionality of the networks, which depend on give block size, as a trade-off between model complexity and prediction performance \cite{li2018hybrid}. Given a network structure, the depth of the convolutional layers is the most predominant factor when dealing with variable input sizes. This means that  increasingly complex architectures are needed for larger block sizes, in order to ensure proper generalisation for these blocks which have higher content variety. Such a factor significantly increases  requirements for  inference because of the number of multiple architectures. 

\begin{figure}[!t]
\centering
\includegraphics[width=0.54\textwidth]{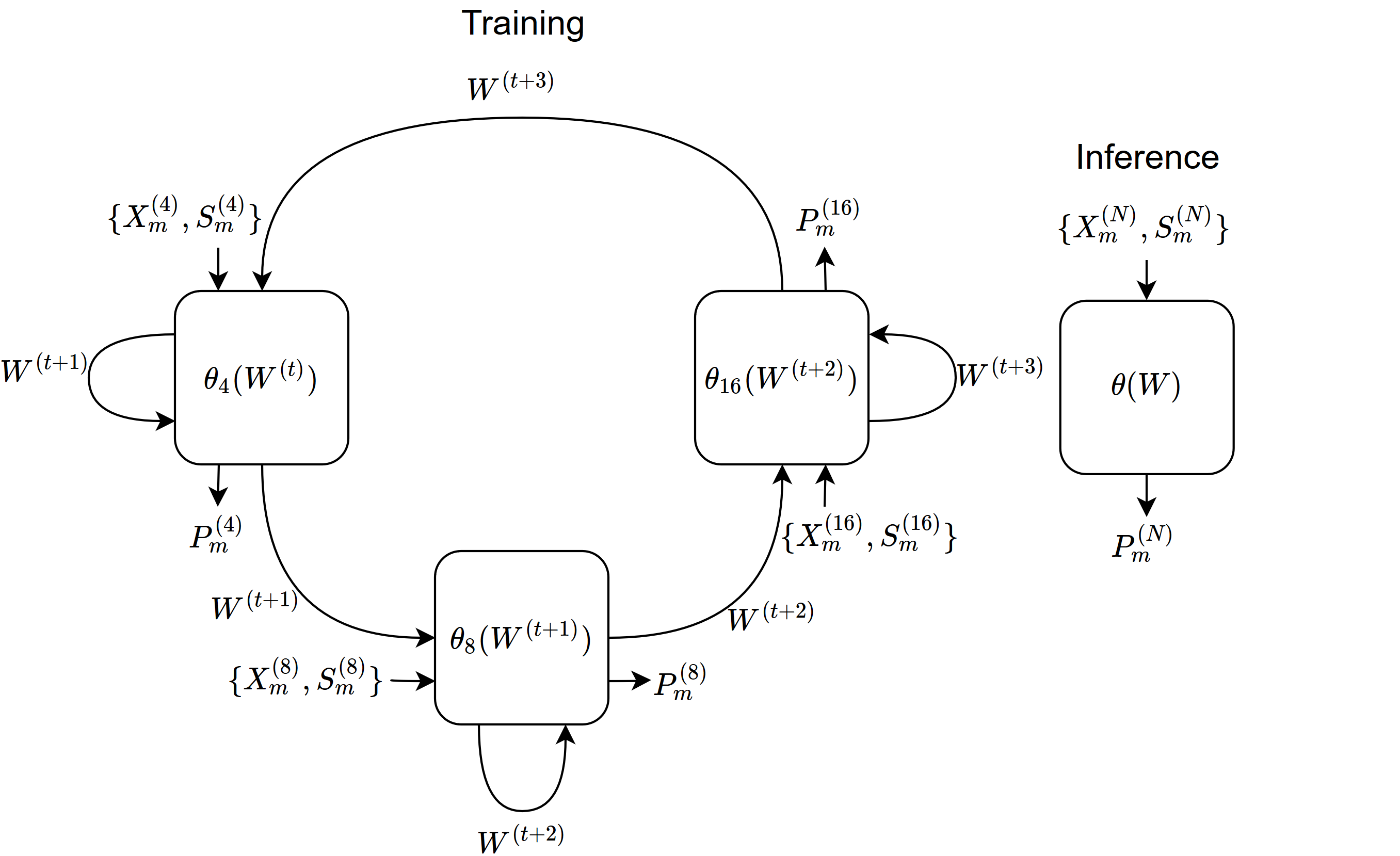}
\caption{Illustration of the proposed multi-model training and inference methodologies. Multiple block-dependent models $\theta_N(W^{(t)})$ are used during training time. A size-agnostic model with a single set of trained weighs $W$ is then used during inference.}
\label{multimodel}
\end{figure}

In order to streamline the inference process, this work proposes a novel multi-model architecture that is independent of the input block size. Theoretically, a convolutional filter can be applied over any input space. Therefore, the fully-convolutional nature of the proposed architecture ($1\times1$ kernels for the cross-component boundary branch and $3\times3$ kernels for the luma convolutional one) allows the design of a size agnostic architecture. As shown in Figure~\ref{multimodel}, the same task can be achieved using multiple models with different input sizes sharing the weights, such that a unified set of filters can be  used a posterior, during inference. The given architecture must employ a number of parameters that is sufficiently large to ensure proper performance for larger blocks, but not too large to incur  overfitting for smaller blocks.

Figure~\ref{train-loop} describes the algorithmic methodology employed to train the multi-model approach. As defined in Section~\ref{sec:attention}, the co-located luma block $X_0\in {\rm I\!R}^{N \times N}$ and the cross-component volume $S_0\in {\rm I\!R}^{3 \times b}$ are considered as inputs to the chroma prediction network. Furthermore, for training of a multi-model the ground-truth is defined as  $Z_{m}^{(N)}$, for a given input $\{X_{m}^{(N)}, S_{m}^{(N)}\}$, and the set of instances from a database of $M$ samples or batches is defined as $\{X_{m}^{(N)}, S_{m}^{(N)}, Z_{m}^{(N)}\}$, where $m = {0, 1 \dots M-1}$ and $N \in $ \{$4$, $8$, $16$\} is the set of supported square block sizes $N \times N$ (the method can be extended to a different set of sizes). As shown in Figure \ref{multimodel}, multiple block-dependent models $\theta_N(W)$ with shared weights $W$ are updated in a concurrent way following the order of supported block sizes. At training step $t$, the individual model $\theta_N(W^{(t)})$ is updated obtaining a new set of weights $W^{(t+1)}$. Finally, a single set of trained weights $W$ is used during inference, obtaining a size-agnostic model $\theta(W)$. Model parameters are updated by minimising the Mean Square Error (MSE) regression loss $\mathcal{L}_{reg}$, as in:
\begin{equation}
\mathcal{L}_{reg}^{(t)} = \frac{1}{C \cdot N^2}\| Z_{m}^{(N)} - \theta_N(X_{m}^{(N)}, S_{m}^{(N)}; W^{(t-1)}\|_{2}^{2},
\label{eq2}
\end{equation}
\noindent where $C=2$ refers to  the number of predicted chroma components, and $\theta_N(W^{(t-1)})$ is the block-dependent model at training step $t-1$.

\begin{figure}  
\begin{algorithmic}[1]  
\Require \{$X_{m}^{(N)}$, $S_{m}^{(N)}$, $Z_{m}^{(N)}$\}, $m \in [0, M)$, $N \in \{4, 8, 16\}$
\Require $\theta_N(W^{(t)})$: $N$ model with shared weights $W^{(t)}$
\Require $\mathcal{L}_{reg}^{(t)}$: Objective function at training step $t$
\State $t\gets 0$ (initialise timestep)
\While{$\theta_{t}$ not converged}
    \For{$m \in [0, M)$}
        \For{$N \in \{4, 8, 16\}$}
            \State $t\gets t+1$
            \State $\mathcal{L}_{reg}^{(t)} \gets MSE(Z_{m}^{(N)},\theta_N(X_{m}^{(N)}, S_{m}^{(N)}; W^{(t-1)}))$
            \State $g^{(t)} \gets \nabla_W \mathcal{L}_{reg}^{(t)}$ (get gradients at step $t$)
            \State $W^{(t)} \gets optimiser(g^{(t)})$
        \EndFor
    \EndFor  
\EndWhile
\end{algorithmic}  
\caption{Training algorithm for the proposed multi-model architecture.}  
\label{train-loop} 
\end{figure}

\subsection{Simplified convolutions}
\label{subsec:int-conv}

Convolutional layers are responsible for most of the network’s complexity. For instance, based on the network hyperparameters from experiments in Section~\ref{sec:experiments}, the luma convolutional branch and the prediction head branch (with $3\times3$ convolutional kernels) alone contain $46,882$ out of $51,714$ parameters, which constitute more than  90\% of the parameters in the entire model. Therefore, the model complexity can be significantly reduced if convolutional layers can be simplified. This subsection explains how a new simplified structure beneficial for practical implementation can be devised by removing activation functions, i.e. by removing non-linearities. It is important to stress that such process is devised only for application on carefully selected layers, i.e. for branches where such simplification does not significantly reduce expected performance.

\begin{figure}[!t]
\centering
\includegraphics[width=0.50\textwidth]{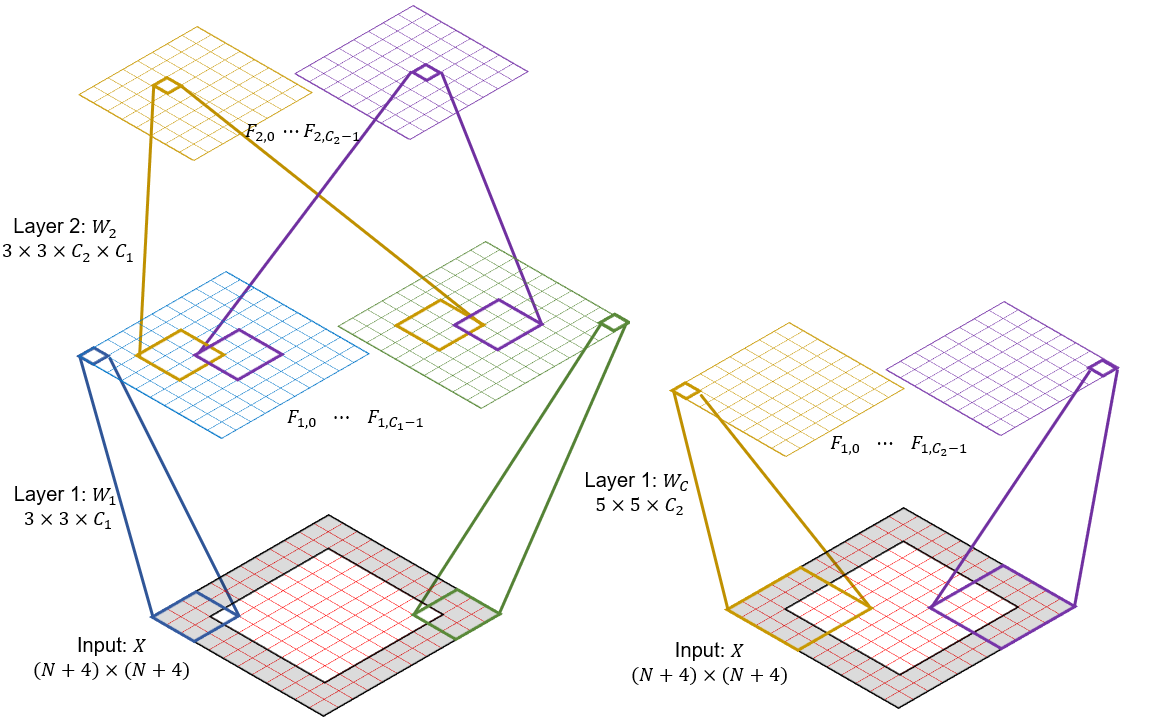}
\caption{Visualisation of the receptive field of a 2-layer convolutional branch with $3\times3$ kernels. Observe that an output pixel in layer $2$ is computed by applying a $3\times 3$ kernel over a field $F_1$ of $3 \times 3$ samples from the first layer's output space. Similarly, each of the $F_1$ values are computed by means of another $3\times 3$ kernel looking at a field $F_0$ of $5\times5$ samples over the input.}
\label{receptive-field}
\end{figure}

Consider specific two-layer convolutional branch (e.g. luma convolutional branch from Figure 2) formulated as:
\begin{equation}
Y =\mathcal{R}(W_2 *\mathcal{R}(W_1 * X + b_1 )+b_2 )
\label{eq3}
\end{equation}
where $C_i$ are the number of features in layer $i$, $b_i \in {\rm I\!R}^{C_i}$ are biases, $K_i \times K_i$ are square convolutional kernel sizes, $W_1 \in {\rm I\!R}^{K_1^2 \times C_0 \times C_1}$ and $W_2 \in {\rm I\!R}^{K_2^2 \times C_1 \times C_2}$ are the weights and bias of the first ($i=1$) and the second ($i=2$) layers, respectively, $C_0$ the dimensions of the input feature map, $\mathcal{R}$ is a Rectified Linear Unit (ReLU) non-linear activation function and $*$ denotes convolution operation. Input to the branch is $X \in {\rm I\!R}^{N^2 \times C_0}$ and the result is a volume of features $Y \in {\rm I\!R}^{N^2 \times C_2}$, which correspond to $X_0$ and $X_2$ from Figure~\ref{icip-arch}, respectively. Removing non-linearities, the given branch can be simplified as:
\begin{equation}
\hat{Y} = W_2 * (W_1 * X + b_1) + b_2,
\label{eq4}
\end{equation}
where it can be observed that a new convolution and bias terms can be defined using trained parameters from the two initial layers, to form a new single layer:
\begin{equation}
\hat{Y} = W_c * X + b_c,
\label{eq5}
\end{equation}
where $W_c \in {\rm I\!R}^{[\hat{K}^2 \times C_0] \times C_2}$ is the function of $W_1$ and $W_2$ with $\hat{K} = K_1 + K_2 - 1$, and $b_c$ is a constant vector derived from $W_2$, $b_1$ and $b_2$. Figure \ref{receptive-field} (a) illustrates the operations performed in Eq. \ref{eq4} for $K_1 = K_2 = 3$ and $C=1$. Analysing the receptive field of the whole branch, a pixel within the output volume $Y$ is computed by applying a $K_2 \times K_2$ kernel over a field $F_1$ from the first layer’s output space. Similarly, each of the $F_1$ values are computed by means of another $K_1 \times K_1$ kernel looking at a field $F_0$. Without the non-linearities, and equivalent of this process is simplified, Figure \ref{receptive-field} (b) and Eq. \ref{eq5}. Notice that $\hat{K} = K_1 + K_2 - 1$ equals $5$ in the example in Figure \ref{receptive-field}. For a variety of parameters, including the values of $C_0$, $C_i$ and $K_i$ used in \cite{grriz2020chroma} and in this paper, this simplification of concatenated convolutional layers allows reduction of model’s parameters at inference time, which will be shown in Section~\ref{sec:architecture_conf}. 

Finally, it should be noted that we limit the removal of activation functions only to branches which include more than one layer, from which at least one layer has $K_i > 1$, and only the activation functions between layers in the same branch are removed (to be able to merge them as in Equation~\ref{eq5}). In such branches with at least one $K_i > 1$ the number of parameters is typically very high, while the removal of non-linearities typically does not impact prediction performance. Activation functions are not removed from the remaining layers. It should be noted that in the attention module and at the intersections of various branches the activation functions are critical and therefore are left unchanged. Section \ref{sec:architecture_conf} performs an ablation test to evaluate the effect of removing the non-linearities, and a test to evaluate how would a convolutional branch directly trained with large-support kernels $\hat{K}$ perform.

\subsection{Simplified cross-component boundary branch}
\label{subsec:simple-ccm}

\begin{figure}[!t]
\centering
\includegraphics[width=0.35\textwidth]{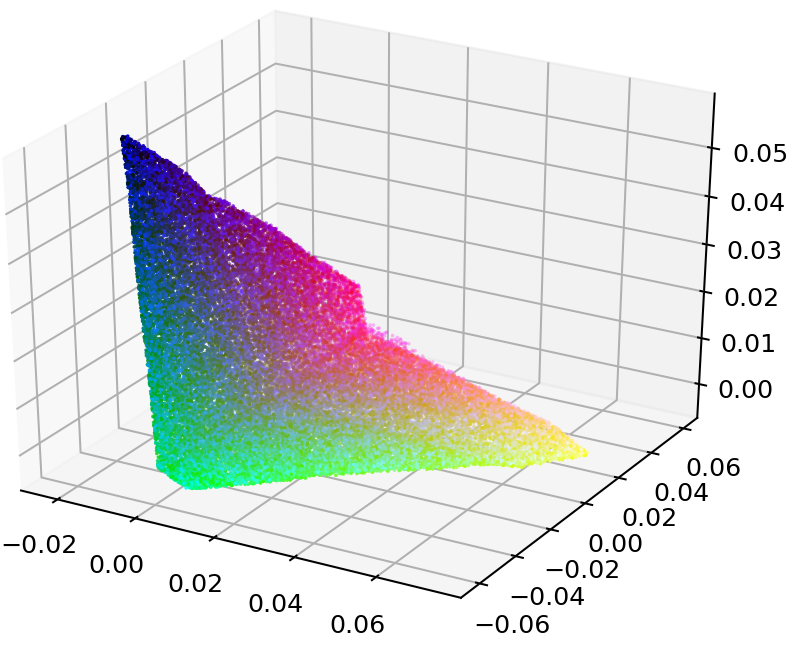}
\caption{Visualisation of the learnt colour space resulting of encoding input YCbCr colours to the 3-dimensional hidden space of the autoencoder.}
\label{colourspace}
\end{figure}

In the baseline model, the cross-component boundary branch transforms the boundary inputs $S\in {\rm I\!R}^{3 \times b}$ into $D_J$-dimensional feature vectors. More specifically, after applying $J=2$ consecutive $1\times1$ convolutional layers, the branch encodes each boundary colour into a high dimensional feature space. It should be noted that a colour is typically represented by 3 components, indexed within a system of coordinates (referred to as the colour space). As such, a three-dimensional feature space can be considered as the space with minimum dimensionality  that is still capable of representing colour information. Therefore, this work proposes the use of autoencoders (AE) to reduce the complexity of the cross-component boundary branch, by compacting the $D$-dimensional feature space into a reduced, $3$-dimensional space. An AE tries to learn an approximation to the identity function $h(x) \approx x$ such that the reconstructed output $\hat{x}$ is as close as possible to the input $x$. The hidden layer will have a reduced dimensionality with respect to the input, which also means that the transformation process may introduce some distortion, i.e. the reconstructed output will not be identical to the input. 

An AE consists of two networks, the encoder $f$ which maps the input to the hidden features, and the decoder $g$ which reconstructs the input from the hidden features. Applying this concept, a compressed representation of the input can be obtained by using the encoder part alone, with the goal of reducing the dimensionality of the input vectors. The encoder network automatically learns how to reduce the dimensions of the input vectors, in a similar fashion to what could be obtained applying a manual Principal Component Analysis (PCA) transformation. The transformation learned by the AE can be trained using the same loss function that is used in the PCA process \cite{bengio2013representation}. Figure~\ref{colourspace} shows the mapping function of the resulting colour space when applying the encoder network over the YCbCr colour space.

Overall, the proposed simplified cross-component boundary branch consists of two $1\times1$ convolutional layers using Leaky ReLU activation functions with a slope $\alpha = 0.2$. First, a $D$-dimensional layer is applied over the boundary inputs $S$ to obtain $S_1 \in {\rm I\!R}^{D \times b}$ feature maps. Then, $S_1$ is fed to the AE's encoder layer $f$ with output $3$ dimensions, to obtain the hidden feature maps $S_2 \in {\rm I\!R}^{3 \times b}$. Finally, a third $1\times1$ convolutional layer (corresponding to the AE decoder layer $g$) is applied to generate the reconstructed maps $\tilde{S}_1$ with  $D$-dimensions. Notice that the decoder layer is only necessary during the training stage to obtain the reconstructed inputs necessary to derive the values of the loss function. Only the encoder layer is needed when using the network, in order to transform the input feature vectors into the $3$ dimensional, reduced vectors. Figure~\ref{schemes} illustrates the branch architecture and its integration within the simplified multi-model. 

Finally, in order to interpret the behaviour of the branch and to identify prediction patterns, a sparsity constraint can be imposed on the loss function. Formally, the following can be used:
\begin{equation}
\mathcal{L}_{AE} = \frac{\lambda_r}{D \cdot b}\|S_1 - \tilde{S}_1\|_{2}^{2} + \frac{\lambda_s}{3 \cdot b}\|S_2\|_{1},
\label{eq7}
\end{equation}
\noindent where the right-most term is used to keep the activation functions in the hidden space remain inactive most of the time, and only return non-zero values for the most descriptive samples. In order to evaluate the effect of the sparsity term, Section \ref{sec:architecture_conf} performs an ablation test that shows its positive regularisation properties during training.

The objective function in Equation~\ref{eq2} can be updated such that the global multi-model loss $\mathcal{L}$ considers both $\mathcal{L}_{reg}$ and $\mathcal{L}_{AE}$ as:
\begin{equation}
\mathcal{L} = \lambda_{reg}\mathcal{L}_{reg} + \lambda_{AE}\mathcal{L}_{AE}
\label{eq8}
\end{equation}
\noindent where $\lambda_{reg}$ and $\lambda_{AE}$ control the contribution of both losses.

\subsection{Integer precision approximation}
\label{subsec:integers}
While the training algorithm results in IEEE-754 64-bit floating point weights and prediction buffers, an additional simplification is proposed in this paper whereby the network weights and prediction buffers are represented using fixed-point integer arithmetic. This is beneficial for deployment of resulting multi-models in efficient hardware implementations, which complex operations such as Leaky ReLU and softmax activation functions can become serious bottlenecks. All the network weights obtained after the training stage are therefore appropriately quantised to fit 32-bit signed integer values. it should be noted that  integer approximation introduces quantisation errors, which may have an impact on the performance of the overall predictions. 

In order to prevent arithmetic overflows after performing multiplications or additions, appropriate scaling factors are defined for each layer during each of the network prediction steps. To further reduce the complexity of the integer approximation, the scaling factor $K_l$ for a given layer $l$ is obtained as a power of $2$, namely $K_l=2^{O_{l}}$, where $O_l$ is the respective precision offset. This ensures that  multiplications can be performed by means of simple binary shifts. Formally, the integer weights $\tilde{W}_l$ and biases $\tilde{b}_l$ for each layer $l$ in the network with weights $W_l$ and bias $b_l$ can be obtained as:
\begin{equation}
\tilde{W}_l = \floor{W_l \cdot 2^{O_l}}; 
\tilde{b}_l = \floor{b_l \cdot 2^{O_l}} .
\label{eq9}
\end{equation}
The offset $O_l$ depends on the offset used on the previous layer $O_{l-1}$, as well as on an internal offset $O_x$ necessary to preserve as much decimal information as possible, compensating for the quantisation that occurred in the previous layer, namely  $O_l = O_x - O_{l-1}$. 

Furthermore, in this approach the values predicted by the network are also integers. In order to avoid defining large internal offsets at each layer, namely having large values of $O_x$, an additional stage of compensation is applied to the predicted values, to keep their values in the range of 32-bit signed integer. For this purpose, another offset $O_y$ is defined, computed as $O_y = O_x - O_l$. The values generated by layer $l$ are then computed as:
\begin{equation}
Y_l = ((\tilde{W}_{l}^T X_{l} + \tilde{b}_l) + (1 << (O_y - 1)) ) >> O_y,
\end{equation}
\noindent where $<<$ and $>>$ represent the left and right binary shifts, respectively, and the offset $(1 << (O_y - 1))$ is considered to reduce the rounding error.

Complex operations requiring floating point divisions need to be approximated to integer precision. The Leaky ReLU activation functions applied on the cross-component boundary branch use a slope $\alpha = 0.2$ which multiplies the negative values. Such an operation can be simply approximated by defining a new activation function $\tilde{A}(x)$ for any input $x$ as follows:
\begin{equation}
\tilde{A}(x) = \left\{
  \begin{array}{lr}
    0 & : x \ge 0\\
    26 \cdot x >> 7 & : x < 0
  \end{array}
\right\}
\label{eq11}
\end{equation}

Conversely, the softmax operations used in the attention module are approximated following a more complex methodology, similar to the one used in \cite{geng2018hardware}. Consider the matrix $M$ as defined in Equation~\ref{eq1} and a given row $j$ in $M$, and a vector $m_j$ as input to the softmax operation. First, all elements $m_j$ in a row are subtracted by the maximum element in the row, namely:
\begin{equation}
\hat{m}_{i, j} = (m_{i, j}/T - max_{\substack{i}}(m_{i,j}/T))
\label{eq12new}
\end{equation}
\noindent where $T$ is the temperature of the softmax operation, set to $0.5$ as previously mentioned. The transformed elements $\hat{m}_{i, j}$ range between the minimum signed integer value and zero, because the arguments $\hat{m}_{i, j}$ are obtained by subtracting the elements in $M$ by the maximum element in each row. To further reduce the possibility of overflows, this range is further clipped to a minimum negative value, set to pre-determined number $V_e$, so that any $\hat{m}_{i, j} < V_e$ is set equal to $V_e$. 

The elements $\hat{m}_{i, j}$ are negative integer numbers within the range $[V_e, 0]$, meaning there is a fixed number of $N_e = \lvert V_e\rvert + 1$ possible values they can assume. To further simplify the process, such an exponential function is replaced by a pre-computed look-up table containing  $N_e$ integer elements. To minimise the approximation error, the exponentials are scaled by a given scaling factor before being approximated to the nearest integer and stored in the corresponding look-up table $\text{\textit{LUT-EXP}}$. Formally, for a given index $k$, where $0 \leq k \leq N_e - 1$, the $k$-th integer input is obtained as $s_k = V_e + k$. The $k$-th element in the look-up table can then be computed as the approximated, scaled exponential value for $s_k$, or:
\begin{equation}
\text{\textit{LUT-EXP}}(k) =\floor{K_{e}e^{s_k}}
\label{eq13}
\end{equation}
\noindent where $K_{e} = 2^{O_{e}}$ is the scaling factor, chosen in a way to maximise the preservation of the original decimal information. When using the look-up table during the prediction process, given an element $\hat{m}_{i, j}$ the corresponding index $k$ can be retrieved as: $k~=~\lvert V_e - \hat{m}_{i, j}\rvert$, to produce the numerator in the softmax function.

The integer approximation of the softmax function can then be written as:
\begin{equation}
\hat{\alpha}_{j, i} = \frac{\text{\textit{LUT-EXP}}(\lvert V_e - \hat{m}_{i, j}\rvert)}{ D(j)},\label{eq1INTEGER}
\end{equation}

\noindent where:
\begin{equation}
D(j)=  \sum_{n=0}^{{b}-1} \text{\textit{LUT-EXP}}(\lvert V_e - \hat{m}_{n, j}\rvert),\label{eq1DENOMINATOR}
\end{equation}

Equation~\ref{eq1INTEGER} implies performing an integer division between the numerator and denominator. This is not ideal, and integer divisions are typically avoided in low complexity encoder implementations. A simple solution to remove the integer division can be obtained by replacing it with a binary shift. However, a different approach is proposed in this paper to provide a more robust approximation that introduces smaller errors in the division. The denominator $D(j)$ as in Equation~\ref{eq1DENOMINATOR} is obtained as the sum of $b$ values extracted from $\text{\textit{LUT-EXP}}$, where $b$ is the number of reference samples extracted from the boundary of the block. As such, the largest blocks under consideration ($16 \times 16$) will result in the largest possible value of reference samples $b_{MAX}$. This means that the maximum value that this denominator can assume is obtained when $b = b_{MAX}$ and when all input $\hat{m}_{i, j} = 0$ (which correspond to $\text{\textit{LUT-EXP}}(\lvert V_e\rvert) = K_{e}$), corresponding to $V_s = b_{MAX} K_{e}$. Similarly, the minimum value (obtained when $\hat{m}_{i, j} =  V_e$) is $0$. Correspondingly, $D(j)$, can assume any positive integer value in the range $[0, V_s]$.

Considering a given scaling factor $K_{s} = 2^{O_{s}}$,  integer division by $D(j)$ can be approximated using a multiplication by the factor $M(j) = \floor{K_{s} / D(j)}$. A given value of $M(j)$ could be computed for all $V_s + 1$ possible values of $D(j)$. Such values can then be stored in another look-up table $\textit{LUT-SUM}$. Clearly though, $V_s$ is too large which means  $\textit{LUT-SUM}$ would be impractical to use due to storage and complexity constraints. For that reason, a smaller table is used, obtained by quantising the possible values of  $D(j)$. A pre-defined step $Q$ is used, resulting in $N_s = (V_s + 1) / Q$ quantised values of $D(j)$. The table $\textit{LUT-SUM}$ of size $N_s$ is then filled accordingly, where each element in the table is obtained as:
\begin{equation}
\text{\textit{LUT-SUM}}(l) =\floor{K_{s} / (l Q)}
\label{eq14}
\end{equation}

Finally, when using the table during the prediction process, given an integer sum $D(j)$, the corresponding index $l$ can be retrieved as: $l~=~\floor{D(j) / Q}$. Following from these simplifications, given an input $\hat{m}_{i, j}$ obtained as in Equation~\ref{eq12new}, the integer sum $D(j)$ obtained from Equation~\ref{eq1DENOMINATOR}, and a quantisation step $Q$, the simplified integer approximation of the softmax function can eventually  be obtained as:
\begin{equation}
\tilde{\alpha}_{j, i} = \text{\textit{LUT-EXP}}(\lvert V_e - \hat{m}_{i, j}\rvert) \cdot \text{\textit{LUT-SUM}}(\floor{D(j) / Q}),\label{eq1INTEGERSIMPLIFIED}
\end{equation}

Notice that $\tilde{\alpha}_{j, i}$ values are finally scaled by $K_o = K_e \cdot K_s$. 

\section{Experiments}
\label{sec:experiments}
\subsection{Training settings}
\label{sec:experiment-settings}
Training examples were extracted from the DIV2K dataset \cite{timofte2017ntire}, which contains high-definition high-resolution content of large diversity. This database contains $800$ training samples and $100$ samples for validation, providing $6$ lower resolution versions with downsampling by  factors of $2$, $3$ and $4$ with a bilinear and unknown filters. For each data instance, one resolution was randomly selected and then M blocks of each $N\times N$ sizes ($N=4, 8, 16$) were chosen, making balanced sets between block sizes and uniformed spatial selections within each image. Moreover, 4:2:0 chroma sub-sampling is assumed, where the same downsampling filters implemented in VVC are used to downsample co-located luma blocks to the size of the corresponding chroma block. All the schemes were trained from scratch using the Adam optimiser \cite{kingma2014adam} with a learning rate of $10^{-4}$.

\subsection{Integration into VVC}
The methods introduced in the paper where integrated within a VVC encoder, using the VVC Test Model (VTM) 7.0 \cite{chen2018algorithmvtm7}. The integration of the proposed NN-based cross-component prediction into the VVC coding scheme requires normative changes not only in the prediction process, but also in the way the chroma intra-prediction mode is signalled in the bitstream and parsed by the decoder. 

A new block-level syntax flag is introduced to indicate whether a given block makes use of one of the proposed schemes. If the proposed NN-based method is used, a prediction is computed for the two chroma components. No additional information is signalled related to the chroma intra-prediction mode for the block. Conversely, if the method is not used, the encoder proceeds in signalling the chroma intra-prediction mode as in conventional VVC specifications. For instance, a subsequent flag is signalled to identify if conventional LM modes are used in the current block or not. The prediction path also needs to accommodate the new NN-based predictions. This largely reuses prediction blocks that are needed to perform conventional CCLM modes. In terms of mode selection at the encoder side, the new NN-based mode is added to the conventional list of modes to be tested in full rate-distortion sense.

\subsection{Architecture configurations}
\label{sec:architecture_conf}
The proposed multi-model architectures and simplifications (Section~\ref{sec:simplifications}) are implemented in 3 different schemes:
\begin{itemize}
    \item Scheme 1: Multi-model architecture (Section~\ref{subsec:multi-model}) applying the methodology in Section~\ref{subsec:int-conv} to simplify the convolutional layers within the luma convolutional branch and the prediction branch, as illustrated in Figure~\ref{schemes}.
    \item Scheme 2: The multi-model architecture in Scheme 1 applying the methodology in Section~\ref{subsec:simple-ccm} to simplify the cross-component boundary branch. As shown in Figure~\ref{schemes}, the integration of the simplified branch requires modification of the initial architecture with changes in the attention module and the prediction branch.
    \item Scheme 3: Architecture in Scheme 1 with the integer precision approximations described in Section~\ref{subsec:integers}.
\end{itemize}

\begin{table}
\centering
\caption{Network hyperparameters during training}
\begin{tabular}{|r||c|c|} 
\toprule
\textbf{Branch} ($C_{in}, K \times K, C_{out}$) & \textbf{Scheme 1 \& 3}                                                                            & \textbf{Scheme 2}                                                                            \\ 
\hline
CC Boundary                       & \begin{tabular}[c]{@{}c@{}}$3, 1 \times 1, 32$\\$32, 1 \times 1, 32$\end{tabular}                   & \begin{tabular}[c]{@{}c@{}}$3, 1 \times 1, 32$\\$32, 1 \times 1, 3$\end{tabular}                    \\ 
\hline
Luma Convolutional                & \begin{tabular}[c]{@{}c@{}}$1, 3 \times 3, 64$\\$64, 3 \times 3, 64$\end{tabular}                   & \begin{tabular}[c]{@{}c@{}}$1, 3 \times 3, 64$\\$64, 3 \times 3, 64$\end{tabular}                   \\ 
\hline
Attention Module                  & \begin{tabular}[c]{@{}c@{}}$32, 1 \times 1, 16$\\$64, 1 \times 1, 16$\\$64, 1 \times 1, 32$\end{tabular} & \begin{tabular}[c]{@{}c@{}}$32, 1 \times 1, 16$\\$64, 1 \times 1, 16$\\$64, 1 \times 1, 3$\end{tabular}  \\ 
\hline
Prediction Head                       & \begin{tabular}[c]{@{}c@{}}$32, 3 \times 3, 32$\\$32, 1 \times 1, 2$\end{tabular}                    & \begin{tabular}[c]{@{}c@{}}$3, 3 \times 3, 3$\\$3, 1 \times 1, 2$\end{tabular}                     \\
\bottomrule
\end{tabular}
\label{train-hyper}
\end{table}

\begin{table}
\centering
\caption{Network hyperparameters during inference}
\begin{tabular}{|r||c|c|} 
\toprule
\textbf{Branch} ($C_{in}, K \times K, C_{out}$) & \textbf{Scheme 1 \& 3}                                                                            & \textbf{Scheme 2}                                                                            \\ 
\hline
CC Boundary                       & \begin{tabular}[c]{@{}c@{}}$3, 1 \times 1, 32$\\$32, 1 \times 1, 32$\end{tabular}                   & \begin{tabular}[c]{@{}c@{}}$3, 1 \times 1, 32$\\$32, 1 \times 1, 3$\end{tabular}                    \\ 
\hline
Luma Convolutional                & \begin{tabular}[c]{@{}c@{}}$1, 5 \times 5, 64$\end{tabular}                   & \begin{tabular}[c]{@{}c@{}}$1, 5 \times 5, 64$\end{tabular}                   \\ 
\hline
Attention Module                  & \begin{tabular}[c]{@{}c@{}}$32, 1 \times 1, 16$\\$64, 1 \times 1, 16$\\$64, 1 \times 1, 32$\end{tabular} & \begin{tabular}[c]{@{}c@{}}$32, 1 \times 1, 16$\\$64, 1 \times 1, 16$\\$64, 1 \times 1, 3$\end{tabular}  \\ 
\hline
Prediction Head                       & \begin{tabular}[c]{@{}c@{}}$32, 3 \times 3, 2$\end{tabular}                    & \begin{tabular}[c]{@{}c@{}}$3, 3 \times 3, 2$\end{tabular}                     \\
\bottomrule
\end{tabular}
\label{test-hyper}
\end{table}

In contrast to previous state-of-the-art methods, the proposed multi-model does not need to adapt its architecture to the input block size. Notice that the fully-convolutional architecture introduced in \cite{grriz2020chroma} enables this design and is able to significantly reduce the complexity of the cross-component boundary branch in \cite{li2018hybrid}, which uses size-dependent fully-connected layers. Table~\ref{train-hyper} shows the network hyperparameters of the proposed schemes during training, whereas Table~\ref{test-hyper} shows the resulting hyperparameters for inference after applying the proposed simplifications. As shown in Tables ~\ref{complexity-params} and \ref{psnr-simulation}, the employed number of parameters in the proposed schemes represents the trade-off between complexity and prediction performance, within the order of magnitude of related attention-based CNNs in \cite{grriz2020chroma}. The proposed simplifications significantly reduce (around 90\%) the original training parameters, achieving lighter architectures for inference time. Table ~\ref{complexity-params} show that the inference version of Scheme 2 reduces to around 85\%, 96\% and 99\% the complexity of the hybrid CNN models in \cite{li2018hybrid} and to around 82\%, 96\% and 98\% the complexity of the attention-based models in \cite{grriz2020chroma}, for $4\times4, 8\times8 $ and $16\times16$ input block sizes, respectively. Finally, in order to provide more insights about the computational cost and compare the proposed schemes with the state-of-the-art methods, Table~\ref{complexity-flops} shows the number of floating point operations (FLOPs) for each architecture per block size. The reduction of operations (e.g. additions and matrix multiplications) to arrive to the predictions is one the predominant factors towards the given speedups. Notice the significant reduction of FLOPs for the proposed inference models.

In order to obtain a preliminary evaluation of the proposed schemes and to compare their prediction performance with the state-of-the-art methods, the trained models were tested on the DIV2K validation set (with 100 multi-resolution images) by means of averaged PSNR. Test samples were obtained with the same methodology as  used in Section~\ref{sec:experiment-settings} for generating the training dataset. Notice that this test uses the training version of the proposed schemes. As shown in Table~\ref{psnr-simulation}, the multi-model approach introduced in Scheme 1 improves the attention-based CNNs in \cite{grriz2020chroma} for  $4 \times 4$ and  $8 \times 8$ blocks, while only a small performance drop can be observed for $16 \times 16$ blocks. However, because of using a fixed architecture for all block sizes, the proposed multi-model architecture averages the complexity of the individual models in \cite{grriz2020chroma} (Table~\ref{complexity-params}), slightly increasing the complexity of the $4 \times 4$ model and simplifying the $16 \times 16$ architecture. The complexity reduction in the $16 \times 16$ model leads to a small drop in performance. As can be observed from Table~\ref{psnr-simulation} , the generalisation process induced by the multi-model methodology (\cite{grriz2020chroma} with multi-model, compared to \cite{grriz2020chroma}) can minimise such drop by distilling knowledge from the rest of block sizes, which is especially evident for $8 \times 8$ blocks where a reduced architecture can improve the state-of-the-art performance. 

Finally, the simplifications introduced in Scheme 2 (e.g. the architecture changes required to integrate the modified cross-component boundary branch within the original model) lower the prediction performance of Scheme 1. However, the highly simplified architecture is capable of outperforming the hybrid CNN models in \cite{li2018hybrid}, observing training PSNR improvements of an additional 1.30, 2.21 and 2.31 dB for $4\times4, 8\times8 $ and $16\times16$ input block sizes, respectively. The combination of attention-based architectures with the proposed multi-model methodology (Scheme 1) considerably improves the NN-based chroma intra-prediction methods in \cite{li2018hybrid}, showing training PSNR improvements by additional 1.93, 1.73 and 2.68 dB for the supported block sizes. In Section~\ref{subsec:simulation} it will be shown how this relatively small PSNR differences lead to significant differences in codec performance.

Several ablations were performed in order to evaluate the effects of the proposed simplifications. First, the effect of the multi-model methodology is evaluated by directly converting the models in [4] to the size-agnostic architecture in Scheme 1 but without the simplifications in Section~\ref{subsec:int-conv} (\cite{grriz2020chroma} with multi-model). As can be shown in Table~\ref{psnr-simulation}, such methodology improves the $4\times4$ and  $8\times8 $ models, with special emphasis in the $8\times8$ case where the number of parameters is smaller than in \cite{grriz2020chroma}. Moreover, the removal of non-linearities towards Scheme 1 does not significantly affect the performance, with a negligible PSNR loss of around 0.3 dB (\cite{grriz2020chroma} with multi-model compared with Scheme 1). Secondly, in order to evaluate the simplified convolutions methodology in Section~\ref{subsec:int-conv}, a version of Scheme 1 was trained with single-layer convolutional branches with large support kernels (e.g. instead of training 2 linear layers with $3\times3$ kernels and then combining them into $5\times5$ kernels for inference, training directly a single-layer branch with $5\times5$ kernels). Experimental results show the positive effects of the proposed methodology, observing a significant drop of performance when a single-layer trained branch is applied (Scheme 1 with single layer training compared with Scheme 1). Finally, the effect of the sparse autoencoder of Scheme 2 is evaluated by removing the sparsity term in Equation~\ref{eq8}. As can be observed, the regularisation properties of the sparsity term, i.e. preventing large activations, boosts the generalisation capabilities of the multi-model and slightly increases the prediction performance by around 0.2 dB. (Scheme 2 without sparsity compared with Scheme 2).

\begin{table}
\centering
\caption{Model complexity per block size}
\begin{tabular}{|r||c|c|c|} 
\toprule
\textbf{Model} (parameters)      & \textbf{$4 \times 4$} & \textbf{$8 \times 8$} & \textbf{$16 \times 16$}  \\ 
\hline
Hybrid CNN \cite{li2018hybrid}                   & $24435$               & $96116$               & $369222$                 \\ 
\hline
Attention-based CNN \cite{grriz2020chroma}          &  $21602$              &  $83106$              &  $186146$                \\ 
\hline
Scheme 1 \& 3 (train/inference) & \multicolumn{3}{c|}{$51714 / 7074$}                                      \\ 
\hline
Scheme 2 (train/inference) & \multicolumn{3}{c|}{ $39371 / 3710$}                                     \\
\bottomrule
\end{tabular}
\label{complexity-params}
\end{table}

 \begin{table}
\centering
\caption{Prediction performance per block size}
\begin{tabular}{|r||c|c|c|} 
\toprule
\textbf{Model} (PSNR) & \textbf{$4 \times 4$}                             & \textbf{$8 \times 8$} & \textbf{$16 \times 16$}  \\ 
\hline
Hybrid CNN \cite{li2018hybrid}             & \begin{tabular}[c]{@{}c@{}}$28.61$\\\end{tabular} & $31.47$               & $33.36$                  \\ 
\hline
Attention-based CNN \cite{grriz2020chroma}     &  $30.23$                                          &  $33.13$              &  $36.13$                 \\
\hline
\cite{grriz2020chroma} with multi-model     &  $30.55$                            &  $33.21$              &  $36.05$                 \\
\hline
Scheme 1 single layer training              &  $30.36$                                          &  $33.05$              &  $35.88$                 \\
\hline
Scheme 2 without sparsity                   &  $29.89$                                          &  $32.66$                  &  $35.64$                 \\
\hline
(proposed) Scheme 1                & $30.54$                                           & $33.20$               & $35.99$                  \\ 
\hline
(proposed) Scheme 2                &  $29.91$                                          & $32.68$               & $35.67$                  \\
\bottomrule
\end{tabular}
\label{psnr-simulation}
\end{table}

\subsection{Simulation Results}
\label{subsec:simulation}
The VVC reference software VTM-7.0 is used as our benchmark and our proposed methodology is tested under the Common Test Conditions (CTC) \cite{ctc}, using the suggested all-intra configuration for VVC with a QP of 22, 27, 32 and 37. In order to fully evaluate the performance of the proposed multi-models, the encoder configuration is constrained to support only square blocks of $4\times4, 8\times8 $ and $16\times16$ pixels. A corresponding VVC anchor was generated under these conditions. BD-rate is adopted to evaluate the relative compression efficiency with respect to the latest VVC anchor. Test sequences include 26 video sequences of different resolutions: $3840 \times 2160$ (Class A1 and A2), $1920 \times 1080$ (Class B), $832 \times 480$ (Class C), $416 \times 240$ (Class D), $1280 \times 720$ (Class E) and screen content (Class F). The ``EncT" and ``DecT" are ``Encoding Time" and ``Decoding Time", respectively.

\begin{table}
\centering
\caption{FLOPs per block size}
\begin{tabular}{|r||c|c|c|} 
\toprule
\textbf{Model} (parameters)      & \textbf{$4 \times 4$} & \textbf{$8 \times 8$} & \textbf{$16 \times 16$}  \\ 
\hline
Hybrid CNN \cite{li2018hybrid}                   & $51465$               & $187273$               & $711945$                 \\ 
\hline
Attention-based CNN \cite{grriz2020chroma}          &  $42795$              &  $165451$              &  $186146$                \\ 
\hline
Scheme 1 \& 3 (train/inference) & \multicolumn{3}{c|}{$102859 / 13770$}                                      \\ 
\hline
Scheme 2 (train/inference) & \multicolumn{3}{c|}{ $79103 / 7225$}                                    \\
\bottomrule
\end{tabular}
\label{complexity-flops}
\end{table}

\begin{table}
\centering
\includegraphics[width=0.434\textwidth]{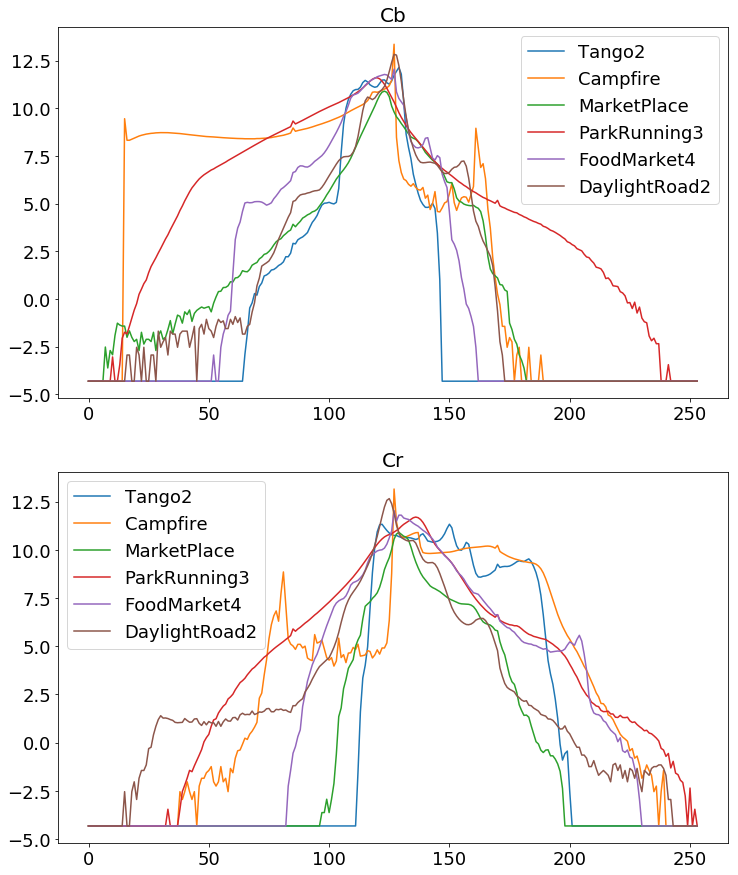}
\captionof{figure}{Comparison of logarithmic colour histograms for different sequences.}
\label{color-analysis}

\vspace{2em}%

\centering

\centering
\captionof{table}{BD-Rates (\%) sorted by Gini index}
\label{table-color-analysis}

\begin{tabular}{|r||c|c|c|c|} 
\toprule
\multirow{2}{*}{Sequence} & \multicolumn{3}{c|}{ \textbf{Scheme 1} } & \multirow{2}{*}{ Gini}  \\ 
\cline{2-4}
                          & Y        & Cb       & Cr                 &                         \\ 
\hline
Tango2                    & ~-0.46~  & ~-8.13~  & ~-3.13~            & ~0.63~                  \\ 
\hline
MarketPlace               & -0.59    & -2.46    & -3.06              & 0.77                    \\ 
\hline
FoodMarket4               & -0.16    & -1.60    & -1.55              & 0.85                    \\ 
\hline
DaylightRoad2             & -0.09    & -5.74    & -1.85              & 0.89                    \\ 
\hline
Campfire                  & -0.21    & 0.14     & -0.88              & 0.98                    \\ 
\hline
ParkRunning3              & -0.31    & -0.73    & -0.77              & 0.99                    \\
\bottomrule
\end{tabular}
\end{table}

A colour analysis is performed in order to evaluate the impact of the chroma channels on the final prediction performance. As suggested in previous colour prediction works \cite{blanch2019end}, standard regression methods for chroma prediction may not be effective for content with wide distributions of colours. A parametric model which is trained to minimise the Euclidean distance between the estimations and the ground truth commonly tends to average the colours of the training examples and hence produce desaturated results. As shown in Figure~\ref{color-analysis}, several CTC sequences are analysed by computing the logarithmic histogram of both chroma components. The width of the logarithmic histograms is compared to the compression performance in Table~\ref {table-color-analysis}. Gini index \cite{davidson2009reliable} is used to quantify the width of the histograms, obtained as
\begin{equation}
Gini(H) = 1 - \sum_{b=0}^{B-1}\left(\frac{H(b)}{\sum_{k=0}^{B-1}H(k)}\right)^2
\label{ginieq}
\end{equation}
\noindent being $H$ a histogram of $B$ bins for a given chroma component. Notice that the average value between both chroma components is used in Table~\ref{table-color-analysis}. A direct correlation between Gini index and coding performance can be observed in Table~\ref{table-color-analysis}, suggesting that Scheme 1 performs better for narrower colour distributions. For instance, the Tango 2 sequence with a Gini index of 0.63 achieves an average Y/Cb/Cr BD-rates of -0.46\%/-8.13\%/-3.13\%, whereas Campfire with wide colour histograms (Gini index of 0.98), obtains average Y/Cb/Cr BD-rates of -0.21\%/0.14\%/-0.88\%. Although the distributions of chroma channels can be a reliable indicator of prediction performance, wide colour distributions may not be the only factor in restricting chroma prediction capabilities of proposed methods, which can be investigated in future work. 

A summary of the component-wise BD-rate results for all the proposed schemes and the related attention-based approach in \cite{grriz2020chroma} is shown in Table~\ref{bd-rate} for all-intra conditions. Scheme 1 achieves an average Y/Cb/Cr BD-rates of -0.25\%/-2.38\%/-1.80\% compared with the anchor, suggesting that the proposed multi-model size agnostic methodology can improve the coding performance of the related attention-based block-dependent models. Besides improving the coding performance, Scheme 1 significantly reduces the encoding (from 212\% to 164\%) and decoding (from 2163\% to 1302\%) times demonstrating the positive effect of the inference simplification. 

Finally, the proposed simplifications introduced in Scheme 2 and Scheme 3 further reduce the encoding and decoding time at the cost of a drop in the coding performance. In particular, the simplified cross-component boundary branch introduced in Scheme 2, achieves an average Y/Cb/Cr BD-rates of -0.13\%/-1.56\%/-1.63\% and, compared to Scheme 1, reduces the encoding (from 164\% to 146\%) and decoding (from 1302\% to 665\%) times. Scheme 3 has lower reduction of encoding time (154\%) than Scheme 2, but it achieves higher reduction in decoding time (665\%), although the integer approximations lowers the performance achieving average Y/Cb/Cr BD-rates of -0.16\%/-1.72\%/-1.38\%.

As described in Section~\ref{sec:simplifications}, the simplified schemes introduced here tackle the complexity reduction of Scheme 1 with two different methodologies. Scheme 2 proposes direct modifications on the original architecture which need to be retrained before being integrated in the prediction pipeline. Conversely, Scheme 3 directly simplifies the final prediction process by approximating the already trained weights from Scheme 1 with integer-precision arithmetic. Therefore, the simulation results suggest that the methodology in Scheme 3 is better at retaining the original performance since a retraining process is not required. However, the highly reduced architecture in Scheme 2 is capable of approximating the performance of Scheme 3 and further reduce the decoder time. 

Overall, the comparison results in Table~\ref{bd-rate} demonstrate that proposed models offer various trade-offs between compression performance and complexity. While it has been shown that the complexity can be significantly reduced, it is still not negligible. Challenges for future work include integerisation of the simplified scheme (Scheme 2) while preventing the compression drop observed for Scheme 3. Recent approaches, including a published one which focuses on intra prediction \cite{schafer2020efficient}, demonstrate that sophisticated integerisation approaches can help retain compression performance of originally trained models while enabling them to become significantly less complex and thus be integrated into future video coding standards.

\begin{table*}
\caption{BD-Rate (\%) of $Y$, $Cb$ and $Cr$ for all proposed schemes and \cite{grriz2020chroma} under all-intra Common Test Conditions}
\begin{tabular}{|c||c|c|c!{\vrule width \lightrulewidth}c|c|c!{\vrule width \lightrulewidth}c|c|c!{\vrule width \lightrulewidth}c|c|c!{\vrule width \lightrulewidth}c|c|} 
\toprule
\multirow{2}{*}{~} & \multicolumn{3}{c!{\vrule width \lightrulewidth}}{Class
  A1} & \multicolumn{3}{c!{\vrule width \lightrulewidth}}{Class
  A2} & \multicolumn{3}{c!{\vrule width \lightrulewidth}}{Class
  B} & \multicolumn{3}{c!{\vrule width \lightrulewidth}}{Class
  C} & \multicolumn{2}{c|}{\multirow{6}{*}{}}                     \\ 
\cline{2-13}
                   & Y              & Cb             & Cr                          & Y              & Cb             & Cr                          & Y              & Cb             & Cr                         & Y              & Cb             & Cr                         & \multicolumn{2}{c|}{}                                      \\ 
\cline{1-13}
Scheme
  1         & \textbf{-0.28} & \textbf{-3.20} & -1.85                       & \textbf{-0.25} & \textbf{-3.11} & \textbf{-1.54}              & \textbf{-0.26} & \textbf{-2.28} & \textbf{-2.33}             & \textbf{-0.30} & \textbf{-1.92} & -1.57                      & \multicolumn{2}{c|}{}                                      \\ 
\cline{1-13}
Scheme
  2         & -0.08          & -1.24          & -1.26                       & -0.12          & -1.59          & -1.31                       & -0.15          & -1.80          & -2.21                      & -0.20          & -1.41          & \textbf{-1.62}             & \multicolumn{2}{c|}{}                                      \\ 
\cline{1-13}
Scheme
  3         & -0.19          & -2.25          & -1.56                       & -0.13          & -2.44          & -1.12                       & -0.16          & -1.78          & -2.05                      & -0.20          & -1.44          & -1.29                      & \multicolumn{2}{c|}{}                                      \\ 
\cline{1-13}
Anchor + \cite{grriz2020chroma}        & -0.26          & -2.17          & \textbf{ -1.96}             & -0.22          & -2.37          & -1.64                       & -0.23          & -2.00          & -2.17                      & -0.26          & -1.64          & -1.41                      & \multicolumn{2}{c|}{}                                      \\ 
\midrule
\multirow{2}{*}{~} & \multicolumn{3}{c!{\vrule width \lightrulewidth}}{Class
  D}  & \multicolumn{3}{c!{\vrule width \lightrulewidth}}{Class E}    & \multicolumn{3}{c!{\vrule width \lightrulewidth}}{Class F}   & \multicolumn{3}{c!{\vrule width \lightrulewidth}}{Overall}   & \multirow{2}{*}{EncT[\%] ~} & \multirow{2}{*}{DecT[\%] ~}  \\ 
\cline{2-13}
                   & Y              & Cb             & Cr                          & Y              & Cb             & Cr                          & Y              & Cb             & Cr                         & Y              & Cb             & Cr                         &                             &                              \\ 
\hline
Scheme
  1         & \textbf{-0.29} & \textbf{-1.70} & \textbf{-1.77}              & \textbf{-0.13} & -1.59          & -1.45                       & \textbf{-0.50} & \textbf{-1.58} & \textbf{-1.99}             & \textbf{-0.25} & \textbf{-2.38} & -1.80                      & 164\%                       & 1302\%                       \\ 
\hline
Scheme
  2         & -0.18          & -1.42          & -1.73                       & -0.08          & \textbf{-1.67} & -1.40                       & -0.34          & -1.50          & -1.90                      & -0.13          & -1.56          & -1.63                      & \textbf{146\%}              & 665\%                        \\ 
\hline
Scheme
  3         & -0.20          & -1.64          & -1.41                       & -0.07          & -0.75          & -0.46                       & -0.37          & -1.24          & -1.43                      & -0.16          & -1.72          & -1.38                      & 154\%                       & \textbf{512\%}               \\ 
\hline
Anchor + \cite{grriz2020chroma}        & -0.25          & -1.55          & -1.67                       & -0.03          & -1.35          & \textbf{ -1.77}             & -0.44          & -1.30          & -1.55                      & -0.21          & -1.90          & \textbf{ -1.81}            & 212\%                       & 2163\%                       \\
\bottomrule
\end{tabular}
\label{bd-rate}
\end{table*}

\section{Conclusion}
\label{sec:conclusions}

This paper showcased the effectiveness of attention-based architectures in performing chroma intra-prediction for video coding. A novel size-agnostic multi-model and its corresponding training methodology were proposed to reduce the inference complexity of previous attention-based approaches. Moreover, the proposed multi-model was proven to better generalise to variable input sizes, outperforming state-of-the-art attention-based models with a fixed and much simpler architecture. Several simplifications were proposed to further reduce the complexity of the original multi-model. First, a framework for reducing the complexity of convolutional operations was introduced and was able to derive an inference model with around 90\% fewer parameters than its relative training version. Furthermore, sparse autoencoders were applied to design a simplified cross-component processing model capable of further reducing the coding complexity of its preceding schemes. Finally, algorithmic insights were proposed to approximate the multi-model schemes in integer-precision arithmetic, which could lead to fast and hardware-aware implementations of complex operations such as softmax and Leaky ReLU activations. 

The proposed schemes were integrated into the VVC anchor VTM-7.0, signalling the prediction methodology as a new chroma intra-prediction mode working in parallel with traditional modes towards  predicting  the  chroma  component  samples. Experimental  results  show  the  effectiveness  of  the  proposed methods, retaining compression efficiency of previously introduced neural network models, while offering 2 different directions for significantly reducing coding complexity, translated to reduced encoding and decoding times. As future work, we aim to implement a complete multi-model for all VVC block sizes  in order to ensure a full usage  of  the  proposed  approach  building on  the  promising results shown in the constrained test conditions. Finally, an improved approach for integer approximations may enable the fusion of all proposed simplifications, leading to a fast and powerful multi-model.
 

%



%
%

\ifCLASSOPTIONcaptionsoff
  \newpage
\fi



\bibliographystyle{IEEEtran}
\bibliography{bibliography}
\end{document}